\documentclass[conference]{IEEEtran}
\IEEEoverridecommandlockouts

\usepackage{xcolor}
\usepackage{breqn}
\usepackage{relsize}
\usepackage{titlesec}

\titlespacing\section{0pt}{4pt plus 4pt minus 4pt}{2pt plus 4pt minus 4pt}
\titlespacing\subsection{0pt}{3pt plus 0pt minus 0pt}{2pt plus 0pt minus 0pt}
\titlespacing\subsubsection{0pt}{12pt plus 4pt minus 2pt}{0pt plus 2pt minus 2pt}

\usepackage{dsfont}
\usepackage{interval}
\usepackage{tabularx}
\usepackage{multirow}
\usepackage{amsmath,etoolbox}

\usepackage[ruled,vlined]{algorithm2e}
\usepackage{cite}
\usepackage{amsmath,amssymb,amsfonts}
\usepackage{algorithmic}
\usepackage{graphicx}
\usepackage{textcomp}
\usepackage{xcolor}
\usepackage{enumitem}
\setlist{nolistsep,leftmargin=.6cm}
\def\BibTeX{{\rm B\kern-.05em{\sc i\kern-.025em b}\kern-.08em
    T\kern-.1667em\lower.7ex\hbox{E}\kern-.125emX}}
\usepackage{amsmath, amssymb, bm, cite, epsfig, psfrag}
\usepackage{graphicx}
\usepackage{cuted}
\usepackage{soul} 
\usepackage[margin=0.7in]{geometry}
\usepackage{dblfloatfix}
\usepackage{array}
\usepackage{lipsum}
\newcolumntype{P}[1]{>{\centering\hspace{0pt}}p{#1}}
\newcolumntype{M}[1]{>{\centering\hspace{0pt}}m{#1}}
\newcolumntype{L}{>{\centering\arraybackslash}m{3cm}}
\usepackage[font=small, skip=0pt]{caption}
\usepackage{epstopdf}	
\usepackage{bbm}
\usepackage{longtable}
\usepackage{supertabular,booktabs}
\usepackage{bbm}
\usepackage{multirow}

\usepackage{subfigure}
\usepackage{etoolbox}
\usepackage{parskip} 
\usepackage{pbox}
\usepackage{fixltx2e}
\usepackage{tabu}	
\usepackage{filecontents}
\usepackage{enumerate}
\usepackage{textcomp}

\usepackage{colortbl}
\usepackage{fancyhdr}

\usepackage{bm}
\newtoggle{conference}
\togglefalse{conference} 
\makeatletter
\newsavebox\myboxA
\newsavebox\myboxB
\newlength\mylenA
\newcommand*\xoverline[2][0.75]{%
    \sbox{\myboxA}{$\m@th#2$}%
    \setbox\myboxB\null
    \ht\myboxB=\ht\myboxA%
    \dp\myboxB=\dp\myboxA%
    \wd\myboxB=#1\wd\myboxA
    \sbox\myboxB{$\m@th\overline{\copy\myboxB}$}
    \setlength\mylenA{\the\wd\myboxA}
    \addtolength\mylenA{-\the\wd\myboxB}%
    \ifdim\wd\myboxB<\wd\myboxA%
       \rlap{\hskip 0.5\mylenA\usebox\myboxB}{\usebox\myboxA}%
    \else
        \hskip -0.5\mylenA\rlap{\usebox\myboxA}{\hskip 0.5\mylenA\usebox\myboxB}%
    \fi}

\begin{document}
\setlength{\abovedisplayskip}{0pt}
\setlength{\belowdisplayskip}{0pt}
\title{ Capacity Bounds and User Identification Costs in Rayleigh-Fading Many-Access Channel}

\author{\IEEEauthorblockN{Jyotish Robin, Elza Erkip}
\IEEEauthorblockA{\textit{Dept. of Electrical and Computer Engineering,}\\
\textit{Tandon School of Engineering, New York University, Brooklyn, NY, USA}}
}

\maketitle
\begin{abstract}
Many-access channel (MnAC) model allows the number of users in the system and the number of active users to scale as a function of the blocklength and as such is suited for dynamic communication systems with massive number of users such as the \textit{Internet of Things}. Existing MnAC models assume a priori knowledge of channel gains which is impractical since acquiring Channel State Information (CSI) for massive number of users can  overwhelm the available radio resources. This paper incorporates Rayleigh fading effects to the MnAC model and derives an upper bound on the symmetric message-length capacity of the Rayleigh-fading Gaussian MnAC. Furthermore, a lower bound on the minimum number of channel uses for discovering the active users is established. In addition, the performance of Noisy-Combinatorial Orthogonal Matching Pursuit (N-COMP) based group testing (GT) is studied as a practical strategy for active device discovery. Simulations show that, for a given SNR, as the number of users increase, the required number of  channel uses for N-COMP GT scales approximately the same way as the lower bound on minimum user identification cost. Moreover, in the low SNR regime, for sufficiently large population sizes, the  number of channel uses required by N-COMP GT was observed to be within a factor of two of the lower bound when the expected number of active users scales sub-linearly with the total population size. 
\end{abstract}

\begin{IEEEkeywords}
  Many-Access channel, Group testing, Active device discovery, multiple-access, Rayleigh-fading
\end{IEEEkeywords}
\vspace{-0.1cm}
\section{Introduction}~\label{sec:intro}
Recent advances in the Internet of Things (IoT)  have lead to envisioning an enormous network of smart devices supporting applications in a wide range of sectors including smart health, industrial automation, smart power systems,  etc. These massive networks are often composed of devices with sparse and sporadic activities. Furthermore, the number of devices that require the channel access can naturally scale as a function of the blocklength \cite{8849781}. However, classical multi-user information theory \cite{10.5555/1146355} assumes a fixed set of users, typically small in number and thus renders itself inadequate for modeling the dynamic nature of massive wireless networks.

Recently, Chen \textit{et.al} \cite{7852531} proposed a novel multiple-access channel model referred to as many-access channel (MnAc) for systems with many transmitters and a single receiver in which the number of transmitters grows with the blocklength $n$. The paper introduced  the notion of `message-length capacity' for a Gaussian MnAC defined as the total number of information bits reliably transmitted in $n$ channel uses. While \cite{7852531} assumes knowledge of each user's channel gain in advance, in a massive wireless communication system, it is often infeasible to acquire the exact Channel State Information (CSI) a priori at the receiver owing to a large number of transmitters. 

In this paper, we consider a fading-MnAc where each user experiences a Rayleigh fading channel whose gain is unknown a priori to the base station. We derive an upper bound on the  message-length capacity and characterize the minimum active user identification cost defined as the minimum number of channel uses to identify the active devices such that the probability of misdetection or false alarms vanishes as the number of users increase asymptotically. We also investigate the performance gap incurred by an on-off keying based group testing (GT) approach for active device discovery in a Rayleigh-fading AWGN channel. 

GT has been previously studied in the context of active device discovery in massive access when there is no noise \cite{8262800,robin2021sparse}. In \cite{6157065}, authors consider several noisy GT scenarios with additive noise as well as dilution noise models. However, these models render themselves insufficient for the Rayleigh fading MnAc we consider in this paper. A related but different problem of neighbor
identification is studied in \cite{5394812}  under Rayleigh fading assumption for finite-user many-access.

In the GT based active device discovery we propose, each device is assigned a binary signature. During each channel use, devices with a ``1" at their corresponding index in the signature transmit an `ON' signal while the others remain silent. The GT based scheme only requires non-coherent energy detection at the receiver. Moreover, GT is known to be efficient  in sparse vector  recovery applications \cite{8926588} which is the case in IoT environments with sparse and sporadic user activities. We derive closed-form expressions for upper bounds on probability of false alarm and misdetection when Noisy-Combinatorial Orthogonal Matching Pursuit based GT (N-COMP GT) is used for active device discovery.

The remainder of this paper is organized as follows. In Section II, we present the system model for Rayleigh fading MnAc. Bounds on message-length capacity and minimum user identification cost for this channel model are established in Section III. In Section IV, we describe active device discovery using GT and derive relevant bounds on probability of misdetection and false alarm for a GT based active device discovery scheme in Rayleigh fading MnAc. Section V compares the performance of GT based active device discovery with the lower bound on minimum user identification cost. Finally, we conclude our paper in Section VI.
\section{System model} 
In \cite{7852531}, Chen et al. studied the symmetric message length capacity of  
Gaussian many-access channel without fading where the number of users $\ell_n$ is tied to block length $n$. Here, we extend the framework to a Rayleigh fading setting. During each channel use, any user can be independently active with probability $\alpha_{n}$. We use $k_n=\alpha_n \ell_n$ to denote the expected number of active users. When active, each user transmits a message from its message set of cardinality $M$. The received symbols over $n$ channel uses is given by,\vspace{0.1cm}
\begin{equation} \vspace{0.15cm}
    \boldsymbol{{y}}=\boldsymbol{\Tilde{S}} \boldsymbol{H} \boldsymbol{{x}} +\boldsymbol{{w}}
    \label{sysmod}
\end{equation} 
where, \small $\boldsymbol{H}=\left(\begin{array}{ccc}
h_{1}  \textbf{I}_{M} & 0 & 0 \\
0 & \ldots & 0 \\
0 & 0 & h_{l_{n}}  \textbf{I}_{M }
\end{array}\right)$ \normalsize $\in \mathbb{C}^{M \ell_{n} \times M \ell_{n}}.$

  The $i^{th}$ diagonal block of $\boldsymbol{H}$  has a value $h_i$ along its diagonal representing the channel coefficient of user $i$. We assume that $h_i$'s are i.i.d with distribution $ \mathcal {C}\mathcal{N}(0,\sigma^2)$ and unknown a priori. Each column of the matrix $\boldsymbol{\tilde{S}} \in \mathbb{C}^{n \times M \ell_{n}}$ represents the $n$-length codeword corresponding to each of the $M$ non-zero messages of the $\ell_n$ users. The codewords associated with each user satisfy the power constraint
$\frac{1}{n} \sum_{i=1}^{n} \tilde{s}_{j i}^{2}(w) \leq P,  \forall j \in \{1,2, \ldots, \ell_{n}\}$.   $\textbf{w}=\left[w_1, \ldots, w_{n}\right] \in \mathbb{C}^{n \times 1}$ is a  vector of independent circularly
symmetric complex Gaussian noise with distribution $\mathcal{C} \mathcal{N}(0,\sigma_w^{2}).$ For each active user, SNR per symbol $(\rho)$ is defined as $\rho:=\frac{P\sigma^2}{\sigma_w^2}.$ In addition, $\boldsymbol{x}$ represents whether a user is active and  which one of the $M$ codewords is transmitted by the active users.
\begin{equation}
    \boldsymbol{{x}}=\left[\begin{array}{r}
{\left[x_{1}\right]_{1 \times M}} \hspace{.2cm}  
\ldots 
 \hspace{.2cm} {\left[x_{l_{n}}\right]_{1 \times M}}
\end{array}\right]^{T} \vspace{.1cm}
\end{equation}
 \vspace{.1cm}where   $\forall k \in \{1,2\ldots, \ell_n\}$, we have the following:
\begin{equation}
\left[x_{k}\right]_{1 \times M}=\left\{\begin{array}{ll}
\boldsymbol{0}_{({1} \times {M})} & \text { w.p. } 1-\alpha_{n} \\
\boldsymbol{e}_{m_{(1 \times M)}} & \text { w.p. } \frac{\alpha_{n}}{M}, \forall m \in \{1, \ldots, M\} 
\end{array}\right. \vspace{0.1cm}
\end{equation}  Therefore, the entropy of $\left[x_{k}\right]_{1 \times M}$, in nats is 
\begin{equation}
\begin{array}{l}
H(\boldsymbol{x})=\ell_{n} H_{2}\left(\alpha_{n}\right)+\alpha_{n} \ln M.
\end{array}
\end{equation} where $H_{2}(.)$ denotes the binary entropy.
For the MnAC channel in (\ref{sysmod}), as defined in \cite{7852531},  a positive non-decreasing function $B(n)$ of blocklength 
$n$ is said to be a symmetric message-length capacity if, for any  $0 < \epsilon < 1$, there exists a sequence of $(\lceil\exp((1-\epsilon)B(n))\rceil,n)$ codes such that the probability $\mathbb{P}(\hat{{{\boldsymbol{x}}}}\neq{{\boldsymbol{x}}})$ vanishes as $n \rightarrow \infty$. Moreover, $(1+\epsilon)B(n)$ needs to be not achievable asymptotically.

While analyzing the user identification costs, we consider $\ell$ users (instead of $\ell_n$) and let other parameters depend on $\ell$. i.e., we use $k_\ell$ and $\alpha_\ell$ instead \vspace{0.05cm} of $k_n$ and $\alpha_n$ respectively. The binary vector $\textbf{b}=\left[b_1,b_2, \ldots, b_{\ell}\right]^{T}\in \mathbb\{0,1\}^{\ell \times 1}$ is used to represent the activity vector. i.e., $b_i \sim Bern(\alpha_\ell), \forall i \in\{1,2,\ldots,\ell\}$. As defined in\cite{7852531}, the minimum user identification cost is said to be $n(\ell)$ if there exists a signature code of length $n_0 = (1 + \epsilon)n(\ell)$ such that the probability of erroneous identification vanishes as $\ell \rightarrow \infty$, whereas the error probability is strictly bounded away from zero if $n_0 = (1 -\epsilon)n(\ell)$.

Our aim is to understand how $B(n)$ or equivalently $\ln M$ scales in the asymptotic regime of $n$ such that the transmitted codewords are successfully decoded at the receiver. Moreover, we are interested in characterizing the minimum user identification cost associated with the Rayleigh fading MnAc.

\section{Bounds to the message-length capacity and minimum user identification cost}\vspace{-0.1cm}
We start by finding an upper bound to the message-length capacity of Rayleigh fading MnAc.

\noindent \textit{Theorem 1: }
For a user set of cardinality $\ell_n$  where each user is active independently with probability $\alpha_{n}$, assuming $\alpha_n$ and $\ell_n$ scale such that $
\ell_{n} e^{-\delta \ell_{n}\alpha_{n}} \rightarrow 0,$ $\forall \delta >0$, the message-length capacity of a Rayleigh-fading MnAC is upper bounded as \vspace{0.2cm}
\begin{equation}
    \ln M \leq n  C_{s u}\left(\rho\right) -\frac{H_{2}\left(\alpha_{n}\right)}{\alpha_{n}}\label{th1}  
\end{equation} where $\rho=\frac{P\sigma^2}{\sigma_w^2}$ and $C_{su}(\rho)$ is the non-coherent capacity of single user Rayleigh fading channel.

{\em Proof:}
Let $C_{su}(\rho)$ denote the non-coherent capacity of a single user Rayleigh fading channel at SNR $ \rho=\frac{P  \sigma^2}{\sigma_w^{2}} $. Then, using cut-set upper bound, the mutual information between $\boldsymbol{{x}}$ and $\boldsymbol{{y}}$  can be bounded as:
\begin{equation}
    I(\boldsymbol{{x}} ; \boldsymbol{{y}}) \leq  nk_n C_{s u}\left(\rho \right).
    \label{Iub}
\end{equation}

Next, we use the upper bound on $H(\boldsymbol{{x}}  \mid \boldsymbol{{y}} )$ derived in \cite{7852531}. The key arguments are presented below.

Define  $\mathcal{B}_m^\ell(\delta,k)=\left\{{\boldsymbol{x}}\in\mathcal{X}_m^\ell:1\le\parallel{{\boldsymbol{x}}}\parallel_0\le(1+\delta)k\right\}$ for  every $\delta\in\ (0,1).$  Since ${{\boldsymbol{x}}}$  is a binary vector with expected support size of $k_n$, the set $\mathcal{B}_M^{l_n}(\delta,k_n)$ has high probability for large $n$. Let ${E}=\mathbbm{1}\{\hat{{{\boldsymbol{x}}}}\neq{{\boldsymbol{x}}}\}$ indicate the error event, where $\hat{{{\boldsymbol{x}}}}$ is the estimation of ${{\boldsymbol{x}}}$ at the receiver. Thus, as established in \cite{7852531}, we can write
\begin{equation}
H(\boldsymbol{{x}}  \mid \boldsymbol{{y}} ) \leq 2 \ln 2+H(\boldsymbol{{x}}  \mid E, \mathbbm{1}\left\{\boldsymbol{{x}} \in \mathcal{B}_{M}^{\ell_{n}}\left(\delta, k_{n}\right)\right\}, \boldsymbol{{y}}  ).
\end{equation}
Furthermore, if $
\ell_{n} e^{-\delta k_{n}} \rightarrow 0
$, then for large values of $n$,
\begin{align}
    H(\boldsymbol{{x}}  \mid E, \mathbbm{1}\left\{\boldsymbol{{x}} \in \mathcal{B}_{M}^{\ell_{n}}\left(\delta, k_{n}\right)\right\}, \boldsymbol{{y}}  ) \leq 4 \mathrm{P}_{e}^{(n)}(k_{n} \ln M+ \hspace{1cm} \nonumber&\\k_{n}+\ell_{n} H_{2}(\alpha_{n}))+\ln M. 
\end{align} where $\mathrm{P}_{e}^{(n)}$ := $\mathbb{P}(\hat{{{\boldsymbol{x}}}}\neq{{\boldsymbol{x}}})$ is the average probability of error. Thus, \begin{equation}
    H(\boldsymbol{{x}}  \mid \boldsymbol{{y}} ) \leq 4 \mathrm{P}_{e}^{(n)}\left(k_{n} \ln M+k_{n}+\ell_{n} H_{2}\left(\alpha_{n}\right)\right) +\ln 4M.
\label{entroub}
\end{equation}

 Using (\ref{Iub}) and (\ref{entroub}) in $H(\boldsymbol{{x}} )=H(\boldsymbol{{x}}  \mid \boldsymbol{{y}} )+I(\boldsymbol{{x}}  ; \boldsymbol{{y}})$ ,we get \vspace{0.2cm}
 \begin{align}
     \left(1-4 \mathrm{P}_{e}^{(n)}\right) \ln M-\frac{1}{k_{n}} \ln M+\left(1-4 \mathrm{P}_{e}^{(n)}\right) \frac{H_{2}\left(\alpha_{n}\right)}{\alpha_{n}} \nonumber &\\[-0.5ex]\leq n  C_{s u}\left(\rho \right)+\frac{\ln 4}{k_{n}}+4 \mathrm{P}_{e}^{(n)}. \hspace{1cm}
 \end{align}
As $k_n\rightarrow\ \infty$, we have
\begin{equation}
  \Big(1-4 \mathrm{P}_{e}^{(n)}-\frac{1}{k_{n}}\Big)\left(\ln M+\frac{H_{2}\left(\alpha_{n}\right)}{\alpha_{n}}\right)\leq n C_{s u}\left(\rho \right)+\delta+4 \mathrm{P}_{e}^{(n)}  .
 \label{l11}
\end{equation} for any arbitrarily small $\delta >0$. Since $\mathrm{P}_{e}^{(n)}$  vanishes as $k_n\rightarrow\ \infty$, we have,\vspace{0.1cm}
\begin{equation}
\ln M+\frac{H_{2}\left(\alpha_{n}\right)}{\alpha_{n}} \leq(1+\varepsilon) n  C_{s u}\left(\rho \right) \text { for any } \varepsilon>0
\end{equation}

Note that, for SNR values less than $10^{-2}$, it has been proven in \cite{4594960} that the capacity of a single user  Rayleigh  fading  channel  without CSI is approximately given by \vspace{0.2cm} \begin{equation}
    C_{su}\left(\rho \right)=\rho- \frac{\rho\ln \left(1+x_{1}^{2}\right)}{x_{1}^{2}}-\frac{\pi\rho \csc \left(\frac{\pi}{x_{1}^{2}}\right)\left(\frac{\rho}{x_{1}^{2}+x_{1}^{4}}\right)^{\frac{1}{x_{1}^{2}}}}{1+x_{1}^{2}}
    \label{csueq}
\end{equation} \normalsize where $x_1$ is obtained by solving \small
    \begin{align}
        x_{1}^{2}-\left(1+x_{1}^{2}\right) \ln \left(1+x_{1}^{2}\right)-\pi\left(\frac{\rho}{x_{1}^{2}+x_{1}^{4}}\right)^{\frac{1}{x_{1}^{2}}} \csc \left(\frac{\pi}{x_{1}^{2}}\right)\times \nonumber &\\
        \left[1+x_{1}^{2}-\pi \cot \left(\frac{\pi}{x_{1}^{2}}\right)+\ln \left(\frac{\rho}{x_{1}^{2}+x_{1}^{4}}\right)\right]=0
    \end{align} \normalsize 
    \begin{figure}[htb]   
	\centering
	\includegraphics[width=3.1in,height=56mm]{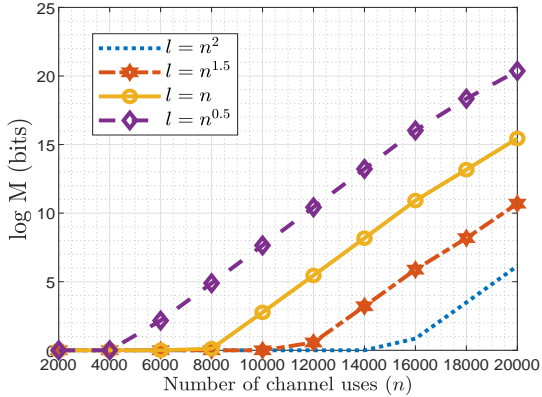}
	\caption{Upper bound on Symmetric message-length capacity of Rayleigh-fading AWGN MnAc at SNR = $10^{-4}$. The average number of active users is assumed to be $k_n=\sqrt{\ell_n}$. } 
	\label{fig:Capacity}
	\end{figure}

Fig. \ref{fig:Capacity} demonstrates the variation in the symmetric message-length capacity (in bits) as a function of the number of channel uses, $n$. We assumed an SNR of $\rho=10^{-4}$ so that  $C_{s u}(\rho)$ can be explicitly computed using (\ref{csueq}).  Here, we considered values of $n$ spanning from 2000 to 20000. For different $\ell_{n}$, we assumed an appropriate $\alpha_n$ such that  $k_n =\sqrt{\ell_{n}}$. For each curve, one can note an initial inertia around a message-length capacity of zero bits. This is because of the active device discovery phase exhausting all the available channel uses.

Now, we will characterize the minimum cost of user identification in terms of the number of channel uses required.

\noindent \textit{Theorem 2:}
For a user set of cardinality $\ell$  where each user is active independently with probability $\alpha_{\ell}$, assuming $\alpha_\ell$ and $\ell$ scale such that $
\ell e^{-\delta \ell\alpha_{\ell}} \rightarrow 0,$ $\forall \delta >0$, the number of channel uses $(n(\ell))$ required to identify the active users  is lower bounded as 
\vspace{0.1cm}\begin{equation} 
    n(\ell) \geq \frac{ H_2(\alpha_{\ell})}{\alpha_{\ell}C_{s u}\left(\rho \right)} .
    \label{e2}
\end{equation} where $\rho=\frac{P\sigma^2}{\sigma_w^2}$ and $C_{su}(\rho)$ is the non-coherent capacity of single user Rayleigh fading channel.

{\em Proof:} The minimum user identification cost corresponds to the case when the codebook of all users contain only their unique ID. i.e., $M=1$. Using $M=1$ in (\ref{th1}) completes the proof.

\section{Active device discovery using GT}~\label{Sysmodel} 
In this section, we analyze an activity detection scheme based on GT and characterize its performance in terms of probability of misdetection and probability of false alarm. Here, all of the $n$ channel uses are dedicated for active device discovery and no data transmission is assumed.
\subsection{Model for GT}
 A signature matrix $\mathbf{S}$ is defined as a binary matrix formed by a  set of  $n$-coordinate binary signatures,  $\mathbf{s}_{i} \in \{0,1\}^{n} \text{ where, } i \in \{1,2, \ldots, \ell\}$. i.e., 
\begin{equation}
    \mathbf{S}=\left[\mathbf{s}_{1}, \ldots, \mathbf{s}_{\ell}\right] \in \mathbb\{0,1\}^{n \times \ell}
    \label{GTmatrix}
    \end{equation}
Each entry of the signature $s_{ij}:=\textbf{s}_i(j)$ takes the value 1  with probability $p$ independent of the other entries in the signature matrix indicating if the $i^{th}$ user is a part of the $j^{th}$ channel use. Otherwise, it takes a value 0.  Note that we use $p =\frac{1}{k_\ell+1}$, the optimality of which is discussed in \cite{8926588}.

In the active device discovery phase, users employ on-off keying to transmit the signatures in $\mathbf{S}$ in a time synchronized manner. During each channel use, the received signal is a superposition of the pulses transmitted by  multiple active users with a 1 at the corresponding indices in their signatures. Thus, over $n$ channel uses, the received vector $\textbf{y}\in\mathbb{C}^{n \times 1}$ can be expressed as $\boldsymbol{{y}}=\sqrt{P}\sum_{j=1}^{\ell} h_{j} b_{j}\boldsymbol{{s}}_{j} +\boldsymbol{{w}}$ where  the noise vector ($\boldsymbol{{w}}$), channel gains ($\boldsymbol{{h}}$)  and power constraint ($P$) are as defined in Section II. We have a non-coherent energy detector which makes a binary decision indicating the presence of energy in the received signal. Note that energy detection does not require any CSI. These decisions are represented by a binary  vector ${\boldsymbol{\tilde{y}}}=\left[{\tilde{y}_1},\tilde{y}_2, \ldots, \tilde{y}_{n}\right]$. The decisions are formed by setting ${\tilde{y}_{i}}$ to 1 if the received energy, $E_r=|y_i|^{2}$ during the $i^{th}$ channel use exceeds a threshold $\tau^{2}$ determined using an appropriate probability of error criterion. During  the active device discovery phase, our aim is to find an estimate of the activity vector $\boldsymbol{\hat{b}}=\left[\hat{b}_1,\hat{b}_2, \ldots, \hat{b}_{\ell}\right]$ using ${\boldsymbol{\tilde{y}}}$ with
a minimum number of channel uses such that $ \lim\limits_{\ell\rightarrow \infty} \mathbb{P}(\boldsymbol{b} \neq \boldsymbol{\hat{b}}) = 0 $. This can be thought of as a GT problem, with each channel use corresponding to a group test \cite{8262800}.

\subsection{Decoding strategy for GT Rayleigh-Fading MnAc}\vspace{0.1cm}
In our analysis, we adapt the N-COMP decoding strategy proposed by Chan $\textit{et.al}$\cite{6120391} to the non-coherent fading channel. In \cite{6120391}, the authors assume a binary symmetric noise model where false negatives and false positives occur with the same probability which is not the case in an AWGN channel with Rayleigh fading.

Let $q_1$ denote the probability that during a channel use, the received energy is below the threshold $\tau^2$ conditioned on the event that there is at least one active user which is a part of the channel use. Mathematically, for an arbitrary channel use index $i \in\{1,2, \ldots,n\}$,
\begin{equation}
   q_1=\mathbb{P}\big(E_r < \tau^2 \Big| \sum_{j=1}^{\ell}b_js_j(i)>0\Big) \vspace{0.1cm}
   \label{q1first} 
\end{equation} 
Assuming, w.l.o.g. that the  $m^{th}$ user is active and is a part of the $i^{th}$ channel use, we can rewrite  $q_1$ in (\ref{q1first}) as
\begin{equation}\vspace{0.1cm}
    q_1=\mathbb{P}\bigg(\bigg|\sqrt{P}h_{m}+\sqrt{P}\sum_{j=1, j \neq m}^{\ell} S_{i j} h_{j} b_{j}+w_{i}\bigg| \leq \tau\bigg)
    \label{q1}
\end{equation}\vspace{0.1cm} for any channel use $i \in \{ 1,2, \ldots n\}.$

Define  $Z=\sqrt{P}h_{m}+\sqrt{P}\sum_{j=1, j \neq m}^{\ell} S_{i j} h_{j} b_{j}+w_{i}$ and $Q\sim Bin(\ell-1, p\alpha_{\ell} )$. Here, $Q$ denotes the number of active users included in the $i^{th}$ channel use other than the $m^{th}$ user.  Thus, (\ref{q1}) becomes
\begin{equation}\vspace{0.1cm}
    q_1= \sum_{g=0}^{\ell-1}\mathbb{P}\left(Q=g\right)P\left(|Z| \leq \tau \Big| Q=g\right). \hspace{0.3cm}
\label{q12}
\end{equation}
Note that conditioned on $Q=g$,  $Z$ is distributed as  $ \mathcal{C} \mathcal{N}(0,\sigma^{2}P(1+g)+\sigma_w^2).$ Thus, conditioned on $Q=g$, $|Z|$ follows Rayleigh distribution given by \begin{equation}
   \mathbb{P}\left(|Z| \leq \tau \vert Q=g\right)=1-\exp \left(-\frac{\tau^{2}}{\sigma^{2}P(1+g)+\sigma_w^2}\right).
    \label{rayleigh}
\end{equation} Combining (\ref{rayleigh}) and  ({\ref{q12}), we can write
    \begin{align}
        q_1=1-\sum_{g=0}^{\ell-1}\mathbb{P}\left(Q=g\right)\exp \left(-\frac{\tau^{2}}{\sigma^{2}P(1+g)+\sigma_w^2}\right) \nonumber&\\
        \geq 1-\exp \left(-\frac{\tau^{2}/\sigma_w^2}{(p\alpha_{\ell}(\ell-1)+1) \rho +1}\right)=q_{1_{LB}}. \hspace{.15cm}\label{q1lb}
    \end{align} 
\vspace{0.1cm}
\noindent In (\ref{q1lb}), we used Jensen's inequality to lower-bound $q_1$ since the RHS in (\ref{q1lb}) is a convex function assuming that the threshold, $\tau^{2}$ used in the energy detector  obeys $\sigma^{2}P+\sigma_w^2 \geq \tau^{2}/2.$

Similarly, let $q_2$ denote the probability that during the $i^{th}$ channel use the received energy is above the threshold $\tau^2$ for any  $i \in \{1,2, \ldots, n\}$ conditioned on the event that there is at least one inactive user. W.l.o.g, assume that the $m^{th}$ user is inactive.  Define  $\tilde{Z}=\sqrt{P}\sum_{j=1, j \neq m}^{\ell} S_{i j} h_{j} b_{j}+w_{i}$ and $Q\sim Bin(\ell-1, p\alpha_{\ell})$.  Conditioned on $Q=g$,  $\tilde{Z}$ is distributed as  $ \mathcal{C} \mathcal{N}(0,\sigma^{2}Pg+\sigma_w^2).$  Thus, similar to (\ref{q1lb}), we can write \vspace{0.1cm}
\begin{align}
    q_2=\sum_{g=0}^{\ell-1}P\left(Q=g\right)P\left(|\tilde{Z}| \geq \tau \vert Q=g\right) \hspace{1.0cm}\nonumber&\\
    \leq \exp \left(-\frac{\tau^{2}/\sigma_w^2}{(p\alpha_{\ell}(\ell-1)+1) \rho +1}\right)=q_{2_{UB}}. \hspace{0cm}\label{q22}
\end{align}

\vspace{0.1cm}With N-COMP based GT, a device is declared as active, if the fraction of channel uses in which the received energy exceeds the predefined threshold is above its expected value. Precisely,
let $\mathcal{N}_\ell=\sum_{i=1}^{n}s_\ell(i)$ denote the number of channel uses in which the user $\ell$ is included and $\mathcal{S}_\ell=\sum_{i=1}^{n}s_\ell(i)\tilde{y_i}$ denote the number of these channel uses in which $\tilde{y}_{i}$ is 1. The decision rule for N-COMP \cite{6120391} decoder can be written as: \vspace{0.1cm}
\begin{equation}
    \text{user $\ell$ is active} \Longleftrightarrow \frac{\mathcal{S}_\ell}{\mathcal{N}_\ell} \geq 1-q_1(1+\Delta) \label{rule}\vspace{0.2cm}
\end{equation}
for a specified $\Delta >0$ optimally chosen based on an appropriate probability of error criterion. \vspace{0.1cm}
\subsection{Probability of misdetection $(P_{MD})$}
\vspace{0.1cm}Consider a set of indicator random variables each corresponding to a misdetection event. Specifically, $\forall j \in\{1,2,\ldots, \ell\}$,
\small
\begin{equation}
    E_{miss}^{j}=\left\{\begin{array}{cc}1, & \text { if } \{\hat{b}_{j}=0\} \cap\{b_{j}=1\}  \\ 0, & \text { otherwise }\end{array}\right.\vspace{0.2cm}
\end{equation} \normalsize The probability of misdetection, $P_{MD}$ can be expressed as 
    \begin{align} P_{\text {MD}} &=\mathbb{P}\Big(\bigcup_{j=1}^{\ell}\left\{E_{miss}^{j}=1\right\}\Big) \nonumber \\ & \leq 
     \sum_{j=1}^{\ell}\mathbb{P}\left({b}_j=1\right)  \mathbb{P}\left(\hat{b}_j=0\vert b_j=1\right)     \label{eq:pmd1}
     \end{align} \normalsize
\vspace{0.1cm}Now, using (\ref{eq:pmd1}) and (\ref{rule}),  $P_{MD}$ can be bounded as:\begin{equation} \small \vspace{0.1cm}
    P_{\text {MD}}\leq 
    \alpha_\ell \mathlarger{\sum}_{j=1}^{\ell}\sum_{r=0}^{n}\mathbb{P}\left(\mathcal{N}_j=r\right) \mathbb{P}\left(\mathcal{S}_j<r(1-q_1(1+\Delta))\right) 
    \label{pmdeq1}
\end{equation} \normalsize Using binomial distribution, (\ref{pmdeq1}) can be simplified to get \vspace{-0.2cm}

\begin{equation} \small
   P_{\text {MD}}\leq \alpha_\ell \ell\mathlarger{\sum}_{r=0}^{n} \bigg(\binom{n}{r}p^{r}(1-p)^{n-r} \sum_{u=r'}^{r}\binom{r}{u} q_{1}^{u}(1-q_1)^{r-u}\bigg)
    \label{pmdeq} 
\end{equation}\normalsize
where $r'=rq_1(1+\Delta)$.

For a random variable $X \sim Bin(r,q_1)$, using Chernoff bound, we have the following inequality:\begin{equation}
   \mathbb{P}(X \geq u)= \sum_{u=r'}^{r}\binom{r}{u} q_{1}^{u}(1-q_1)^{r-u} \leq e^{-2rq_1^{2}\Delta^{2}}
   \label{chernoff}\vspace{0.1cm}
\end{equation}

Using   (\ref{q1lb}) and (\ref{chernoff}) in (\ref{pmdeq}) and applying binomial theorem, we get
$ P_{\text {MD}}
     \leq\alpha_\ell \ell\left(1-p+p e^{-2(q_{1_{LB}} \Delta)^{2}}\right)^{n}.$  Now,  assuming  $n_1=\beta_1 \times \frac{H_2(\alpha_\ell)}{\alpha_\ell}$, we have \vspace{0.1cm}\begin{equation}
    P_{\text {MD}}\leq \alpha_\ell  \ell e^{-\beta_1    \frac{H_2(\alpha_{\ell})}{\alpha_\ell} p(1-e^{-2})(q_{1_{LB}} \Delta)^{2}}
\label{pmd3}
\end{equation}
This result is summarized in the following theorem:

\noindent \textit{Theorem 3:}
For a user set of cardinality $\ell$ where each user can be active independently with probability $\alpha_{\ell}$, the probability of misdetection while using N-COMP GT across $n$ channel uses for discovering active users in a Rayleigh fading MnAc with $\rho$ as SNR per user is upper bounded as  \vspace{0.1cm}
\begin{equation}
    P_{MD} \leq \alpha_\ell \ell e^{-\beta_1  \log(\frac{1}{\alpha_\ell})\times p(1-e^{-2})(q_{1_{LB}} \Delta)^{2}} 
\end{equation}where  $\beta_1= n\times\frac{\alpha_{\ell}} {H_2(\alpha_{\ell})}$, $p=\frac{1}{k_\ell+1}$ and $q_{1_{LB}}$ is given in (\ref{q1lb}).

\noindent\textit{Corollary 1:}
The number of channel uses required so that $P_{MD}$ will be less than  $\ell^{-\delta}$ is given by $n_1=\beta_1 \times \frac{H_2(\alpha_\ell)}{\alpha_\ell}$  where $\beta_1 \geq \frac{\ln(2)}{p(1-e^{-2})(q_{1_{LB}}\Delta)^{2}}\times \bigg(\frac{(1+\delta)\ln(\ell)}{\ln(1/\alpha_\ell)}-1\bigg)$. 

\subsection{Probability of false positive $(P_{FP})$}
We can compute the probability of false positive as follows:
\begin{align}
    P_{\text {FP}}\leq \sum_{j=1}^{\ell}P\left({b}_j=0\right)  P\left(\hat{b}_j=1\vert b_j=0\right)\hspace{4.6cm}\nonumber&\\[-1.5ex]
    =(1-\alpha_\ell) \ell\sum_{r=0}^{n}P\left(\mathcal{N}_j=r\right) P\left(\mathcal{S}_{j}\geq r(1-q_1(1+\Delta))\right)\hspace{1.8cm}\nonumber
\end{align} \vspace{-0.2cm}
    \begin{equation}
       \hspace{0.65cm} =(1-\alpha_\ell) \ell\mathlarger{\sum}_{r=0}^{n} \Bigg(P\left(\mathcal{N}_j=r\right) \sum_{u=r''}^{r}\binom{r}{u} q_{2}^{u}(1-q_2)^{r-u}\Bigg)\hspace{1cm} \label{lll2}
    \end{equation} where $P\left(\mathcal{N}_j=r\right)=\binom{n}{r}p^{r}(1-p)^{n-r}$ and $r''=r(1-q_1(1+\Delta)).$

For $X \sim Bin(n,p)$, if $m<n/2$, then as established in \cite{zhang2020nonasymptotic} the distribution function can be lower bounded as \vspace{0.2cm}
\begin{equation}
    P(X \leq m)\geq \frac{\exp (-n \mathrm{D}(m / n \| p))}{\sigma\sqrt{2 \pi}}
\end{equation} where $D(.)$ is the KL-Divergence between the distributions $Bern(p)$ and $Bern(m/n)$. Thus, we can write the following:\vspace{0.1cm}
\begin{align}
    \sum_{u=r(1-q_1(1+\Delta))}^{r}\binom{r}{u} q_{2}^{u}(1-q_2)^{r-u}\leq 1-\frac{e^{ (-r \mathrm{D}(q_2-q_1\Delta) \| q_2))}}{\sqrt{2 \pi rq_1q_2}} \label{l1}
\end{align} Using (\ref{l1}) in (\ref{lll2}) and invoking Binomial theorem, we get \vspace{0.1cm}
\begin{equation}
    P_{\text {FP}}
\leq (1-\alpha_\ell) \ell\Bigg(1-\bigg(1-p+p e^{\frac{(- \mathrm{D}(q_2-q_1\Delta) \| q_2))}{\sqrt{2 \pi q_1q_2}}}\bigg)^{n}\Bigg)
    \label{pfpeq2}
\end{equation}
Now, assuming $n_2=\beta_2 \times \frac{H_2(\alpha_\ell)}{\alpha_\ell}$, we have the following results:

\noindent\textit{Theorem 4:}
For a user set of cardinality $\ell$ where each user can be active independently with probability $\alpha_{\ell}$, the probability of false alarm while using N-COMP GT across $n$ channel uses for discovering active users in a Rayleigh fading MnAc with $\rho$ as SNR per user  is upper bounded as  \vspace{0.1cm}
\begin{equation}
    P_{FP}\leq  \ell(1-\alpha_\ell) \Big(1-e^{-\beta_2  \log (\frac{1}{\alpha_\ell})  p\big(1-\eta\big)}\Big)
\end{equation} where $\eta=\exp\Big({\frac{- \mathrm{D}(q_2-q_1\Delta) \| q_2)}{\sqrt{2 \pi q_1q_2}}}\Big)$ and $p=\frac{1}{k_\ell+1}$. Here $D(.)$ computes the KL-Divergence between two Bernoulli distributions. $q_1$ and $q_2$ are bounded as in (\ref{q1lb}) and (\ref{q22}).

\noindent \textit{Corollary 2:}
The number of channel uses required so that $P_{FP}$ will be less than  $\ell^{-\delta}$ is given by $n_2=\beta_2 \times \frac{H_2(\alpha_\ell)}{\alpha_\ell}$  where $\beta_2 \geq \frac{1}{p(1-\eta )}\times \frac{1}{\log(1/\alpha_\ell)} \times \log\Big(\frac{(1-\alpha_\ell)\ell^{\delta+1}}{(1-\alpha_\ell)\ell^{\delta+1}-1}\Big)$.
\section{Comparison of N-COMP GT and the lower Bound on  minimum user identification cost}
In this section, through numerical simulations, we compare the performance of N-COMP  GT (in terms of number of channel uses required to have  predetermined probability of misdetection and false positive)  with the minimum user identification cost established in Theorem 2. Though our analysis is valid for general SNR values, in our simulations, we focus on the low-SNR regime since a numerically tractable characterization of the capacity of a single user Rayleigh fading channel without CSI is available as given in (\ref{csueq}) for the low-SNR regime. Our analysis can be translated to arbitrary SNRs  by using the non-coherent capacity expression appropriate for the SNR considered. Also note that the low-SNR results presented here are indeed a lower bound for the high SNR scenarios.
 
\begin{figure}[htb]   
	\centering
	\includegraphics[width=3in,height=53mm]{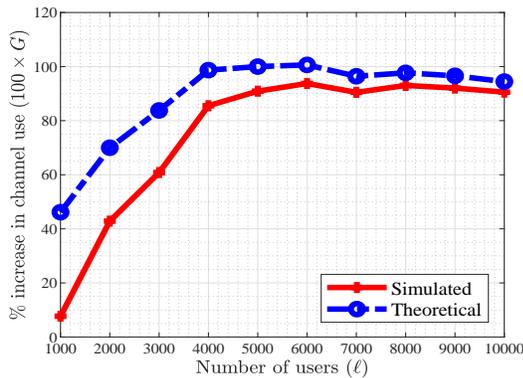}
	\setlength{\belowcaptionskip}{-10pt}
	\caption{	Percentage increase in channel use as a function of $\ell$ at SNR = $10^{-4}$. The average number of active users is assumed to be $k_\ell= \sqrt{\ell}$. } 
	\label{fig:muid}
	\end{figure}

For the probability of misdetection  and probability of false alarm to be less than $\ell^{-\delta}$, the required minimum number of channel uses for N-COMP  GT is \begin{equation} 
 n_{\textsc{GT}} =\max\{\beta_1,\beta_2\} \times \ell H_2(\alpha_\ell)\label{e1}
\end{equation}  
where $\beta_1$ and $\beta_2$ are lower bounded in Corollary 1 and Corollary 2 respectively. Thus, the fractional increase in the number of channel uses N-COMP GT requires w.r.t the minimum user identification cost established in Theorem 2 is given by 
\begin{equation}
    G=\frac{n_{GT}-n_0}{n_0}=\max\{\beta_1,\beta_2\} k_{\ell}C_{s u}\left(\frac{P  \sigma^2}{\sigma_w^{2}} \right)-1.
\end{equation}

	In our numerical analysis, we used the decision rule for N-COMP as in  (26) with  $\Delta = 0.05$ and $p=\frac{1}{k_\ell+1}$. Moreover, we numerically optimized $\tau$ under the constraint $\sigma^{2}P+\sigma_w^2 \geq \tau^{2}/2$ through exhaustive search while simulating (\ref{e1}).
	
\begin{figure}[htb]  
	\centering
	\includegraphics[width=3in,height=53mm]{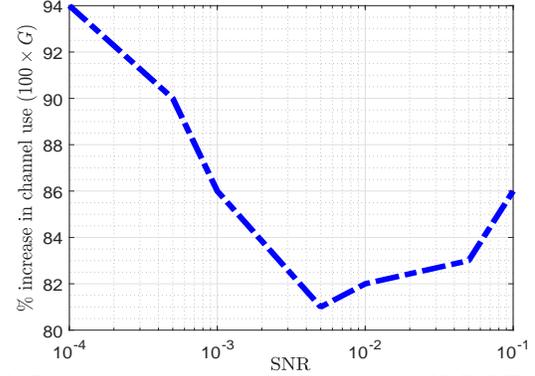}
	\setlength{\belowcaptionskip}{-8pt}
	\caption{	Percentage increase in channel use for N-COMP GT   as a function of SNR when $\ell=10^{4}$. The average number of active users is assumed to be $k_\ell= \sqrt{\ell}$. } 
	\label{fig:excs}
	\end{figure}	
 Fig. \ref{fig:muid} shows the variation in $G$ as a function of number of users $(\ell)$ at an SNR of $10^{-4}$. We have appropriately chosen $\alpha_\ell$ such that the average number of active users is $k_\ell=\sqrt{\ell}$.  We note that the theoretically established upper bound for $n_{\textsc{GT}}$ is tight once the number of users in the system is large.  In Fig. \ref{fig:excs}, how $G$ varies as a function of SNR for a fixed number of users $(\ell=10^{4})$ is studied. One can infer that the number of channel uses required by N-COMP GT is typically within a factor of  two in the low-SNR range. In addition, the slight increase in the performance gap when SNR rises beyond $10^{-2}$ in Fig. \ref{fig:excs} is intuitively due to the lower bound $n(\ell)$ in Theorem 2 being quiet loose at this regime. We also considered several sub-linear scaling  for $k_\ell$ such that $k_{\ell} =\ell^{\gamma}$, where $\gamma \in (0.1,0.5)$ and similar performance (in terms of excess channel uses required) was observed for sufficiently large population sizes.
\section{Conclusion and discussion}
In this paper, we have derived an upper bound on the message-length capacity of a Gaussian many-access channel with Rayleigh Fading assuming no CSI at the receiver. Moreover, we have characterized a lower bound on the minimum number of channel uses required during active device discovery phase. In addition, we have investigated the performance gap incurred by a N-COMP based  GT scheme for active device discovery. Through numerical simulations, it has been verified that in the low SNR regimes, the excess channel uses N-COMP requires is not too far from the lower bound. Thus, in a practical IoT setting, if the number of excess channel uses required does not overburden the system resources, our proposed N-COMP based GT strategy can be useful since it is based on a simple non-coherent energy detection which requires no CSI. Our study can be easily extended to general SNR values by using appropriate expressions for the non-coherent capacity. As a future work, it might be interesting to study how other decoding algorithms for GT can further reduce the performance gap.
\vspace{.3cm}

\vspace{-0.3cm}
\bibliographystyle{IEEEtran}
\bibliography{THz_scatter}

\begin{thebibliography}{10}
\providecommand{\url}[1]{#1}
\csname url@samestyle\endcsname
\providecommand{\newblock}{\relax}
\providecommand{\bibinfo}[2]{#2}
\providecommand{\BIBentrySTDinterwordspacing}{\spaceskip=0pt\relax}
\providecommand{\BIBentryALTinterwordstretchfactor}{4}
\providecommand{\BIBentryALTinterwordspacing}{\spaceskip=\fontdimen2\font plus
\BIBentryALTinterwordstretchfactor\fontdimen3\font minus
  \fontdimen4\font\relax}
\providecommand{\BIBforeignlanguage}[2]{{%
\expandafter\ifx\csname l@#1\endcsname\relax
\typeout{** WARNING: IEEEtran.bst: No hyphenation pattern has been}%
\typeout{** loaded for the language `#1'. Using the pattern for}%
\typeout{** the default language instead.}%
\else
\language=\csname l@#1\endcsname
\fi
#2}}
\providecommand{\BIBdecl}{\relax}
\BIBdecl

\bibitem{8849781}
S.~S. {Kowshik} and Y.~{Polyanskiy}, ``Quasi-static fading {MAC} with many
  users and finite payload,'' in \emph{2019 IEEE International Symposium on
  Information Theory (ISIT)}, 2019, pp. 440--444.

\bibitem{10.5555/1146355}
T.~M. Cover and J.~A. Thomas, \emph{Elements of Information Theory (Wiley
  Series in Telecommunications and Signal Processing)}.\hskip 1em plus 0.5em
  minus 0.4em\relax USA: Wiley-Interscience, 2006.

\bibitem{7852531}
X.~{Chen}, T.~{Chen}, and D.~{Guo}, ``Capacity of {G}aussian many-access
  channels,'' \emph{IEEE Transactions on Information Theory}, vol.~63, no.~6,
  pp. 3516--3539, 2017.

\bibitem{8262800}
H.~A. {Inan}, P.~{Kairouz}, and A.~{Ozgur}, ``Sparse group testing codes for
  low-energy massive random access,'' in \emph{2017 55th Annual Allerton
  Conference on Communication, Control, and Computing (Allerton)}, 2017, pp.
  658--665.

\bibitem{robin2021sparse}
J.~Robin and E.~Erkip, ``Sparse activity discovery in energy constrained
  multi-cluster {IoT} networks using group testing,'' Online Available:
  https://arxiv.org/abs/2103.16174.

\bibitem{6157065}
G.~K. {Atia} and V.~{Saligrama}, ``Boolean compressed sensing and noisy group
  testing,'' \emph{IEEE Transactions on Information Theory}, vol.~58, no.~3,
  pp. 1880--1901, March 2012.

\bibitem{5394812}
J.~{Luo} and D.~{Guo}, ``Compressed neighbor discovery for wireless ad hoc
  networks: The {Rayleigh} fading case,'' in \emph{2009 47th Annual Allerton
  Conference on Communication, Control, and Computing (Allerton)}, 2009, pp.
  308--313.

\bibitem{8926588}
\BIBentryALTinterwordspacing
M.~{Aldridge}, O.~{Johnson}, and J.~{Scarlett}, \emph{Group Testing: An
  Information Theory Perspective}.\hskip 1em plus 0.5em minus 0.4em\relax now,
  2019. [Online]. Available: \url{https://ieeexplore.ieee.org/document/8926588}
\BIBentrySTDinterwordspacing

\bibitem{4594960}
Z.~Rezki, D.~Haccoun, and Gagnon, ``Capacity of the discrete-time non-coherent
  memoryless {G}aussian channels at low {SNR},'' in \emph{2008 IEEE
  International Symposium on Information Theory}, 2008, pp. 121--125.

\bibitem{6120391}
C.~L. {Chan}, P.~H. {Che}, S.~{Jaggi}, and V.~{Saligrama}, ``Non-adaptive
  probabilistic group testing with noisy measurements: Near-optimal bounds with
  efficient algorithms,'' in \emph{2011 49th Annual Allerton Conference on
  Communication, Control, and Computing (Allerton)}, 2011, pp. 1832--1839.

\bibitem{zhang2020nonasymptotic}
A.~R. Zhang and Y.~Zhou, ``On the non-asymptotic and sharp lower tail bounds of
  random variables,'' Online Available: https://arxiv.org/abs/1810.09006.

\end{thebibliography}

\end{document}